# Raman Spectroscopy Insights into the Reorientational Dynamics of Polyanions in Solid Electrolytes

Jianghai Chen,[1#] Yu Yang,[2#] Ya Tang,[3]\* Hong Zhu,[2]\* Wenqian Chen[1]\*

[#]These authors contributed equally to this work

[1] Key Laboratory of Organic Compound Pollution Control Engineering (MOE), School of Environmental and Chemical Engineering, Shanghai University, Shanghai 200444, China.
[2]University of Michigan-Shanghai Jiao Tong University Joint Institute, Shanghai Jiao Tong University, Shanghai 200240, China.
[3] Department of Chemistry, School of Science, Shanghai University, No.99, Shangda Road, Baoshan District, Shanghai, China.

\*Corresponding author: Email: tangya0709@shu.edu.cn (Y.T); hong.zhu@sjtu.edu.cn (H. Z); wenqianchen@shu.edu.cn (WQ. C)

[#] Authors contributed equally.

**Abstract**

The interaction between cation diffusion and polyanion rotational dynamics has garnered significant attention, yet characterizing the rotational dynamics of polyanions remains challenging. Previous studies have primarily relied on complex techniques such as Nuclear Magnetic Resonance (NMR) and Quasi Elastic Neutron Scattering (QENS). In this work, we use *ab initio* molecular dynamics (AIMD) simulations and temperature-dependent Raman spectroscopy to investigate the reorientational dynamics of the $NO_2^-$ polyanion in $Na_3ONO_2$ and $NaNO_2$. Our findings reveal distinct reorientational behaviors and establish a clear correlation between molecular reorientation and Raman spectral features in polyanion-based solid electrolytes. AIMD simulations show that $NO_2^-$ rotates easily in $Na_3ONO_2$, while its rotational dynamics are hindered in $NaNO_2$. Raman spectroscopy confirms these results, with temperature-dependent shifts in the internal vibrational modes of the polyanion. In $Na_3ONO_2$, the full width at half maximum (FWHM) of these modes increases with temperature, while $NaNO_2$ exhibits a constant FWHM until a significant jump at the ordered-disordered phase transition. These insights enhance our understanding of polyanion reorientation and offer a framework for characterizing similar dynamics in other solid electrolytes.

**Introduction**

Advancements in solid-state ion conduction research have significantly contributed to the development of superionic conductors.[1–3] A variety of electrochemical and spectroscopic techniques are employed to investigate ionic conduction across multiple time scales, as shown in Figure 1. Electrochemical Impedance Spectroscopy (EIS)[4,5] is commonly used to study the hopping dynamics within bulk and grain boundaries. Nuclear Magnetic Resonance (NMR)[6–8] and Quasi-Elastic Neutron Scattering (QENS)[9] provide insights into ion coupling dynamics and ion hopping on nanosecond and picosecond time scales, respectively. Recent developments in ultrafast spectroscopy, utilizing terahertz (THz) optical excitation, enables real-time detection of ion hopping on picosecond time scales and allows detailed analysis of complex ion coupling interactions.[10–12] However, these techniques primarily focus on cation hopping, with limited attention given to understanding anion dynamics. In particular, the coupling of phonon-ion vibrations is believed to play a crucial role in the ion conduction mechanism,[13–17] e.g. the polyanion reorientation facilitating cation diffusion. Although QENS and NMR have been employed to detect the reorientation of various polyanions, such as ($PO_4^{3-}$),[18] ($OH^-$),[19,20] ($PS_4^{3-}$),[13,21,22] ($SiS_4^{4-}$),[21] ($YBr_6^{3-}$),[23] and ($YCl_6^{3-}$),[23] their limited time scales and operational complexity have restricted their widespread application.

Polyanion reorientation typically occurs on picosecond to femtosecond time scales,[14,17,24] which fall within the response range of Raman spectroscopy. Raman scattering is an inelastic scattering where incident photons interact with the vibrational modes of a material, resulting in a shift in the photon's energy.[25] This energy shift provides valuable information about the vibrational, rotational, and other low-frequency modes.[26–29] Raman spectroscopy provides several advantages over QENS and NMR for probing material dynamics. Unlike QENS, which requires large-scale equipment and gram-scale samples, Raman spectroscopy permits the detection of milligram-scale samples using cost-effective instruments in standard labs. Although NMR offers high-resolution insights into local structural environments and dynamics, it typically involves long acquisition times and may require isotopic substitutions to achieve sufficient sensitivity. In contrast, Raman spectroscopy is a rapid, non-destructive technique with high sensitivity to vibrational, rotational and low-frequency lattice modes. Its ability to capture temperature-dependent changes in spectral features, such as the shifts and broadening of vibrational peaks, allow researchers to directly correlate these changes with dynamic processes like polyanion reorientation. In 1985, Börjesson et al.[30] demonstrated that the full width at half maximum (FWHM) of internal vibrational modes of a polyanion contains both vibrational and reorientational relaxation broadening. By subtracting the FWHM of the anisotropic spectrum from the isotropic spectrum, they obtained the reorientational broadening of the $SO_4^{2-}$ polyanion in $LiSO_4$ and $LiAgSO_4$ increasing with rising temperature. This observation confirms that the $SO_4^{2-}$ polyanion undergoes reorientation. In addition, since the variation of vibrational broadening with temperature is considerably less pronounced than that of reorientational broadening,[30,31] the contribution of vibrational broadening is often negligible, allowing FWHM variations of internal vibrational modes can be used to infer the polyanion reorientational dynamics. Our previous temperature-dependent Raman analysis confirmed that the $SO_4^{2-}$ polyanion in $Na_3SO_4F$ does not undergo reorientation during its 386 K phase transition.[30] These studies have unveiled the potential of using Raman spectroscopy to investigate the reorientational dynamics of polyanions in solid electrolytes; however, there remains a lack of systematic analyses to guide researchers in understanding the intricate relationship between solid-state Raman and the polyanion reorientation in various electrolyte materials.

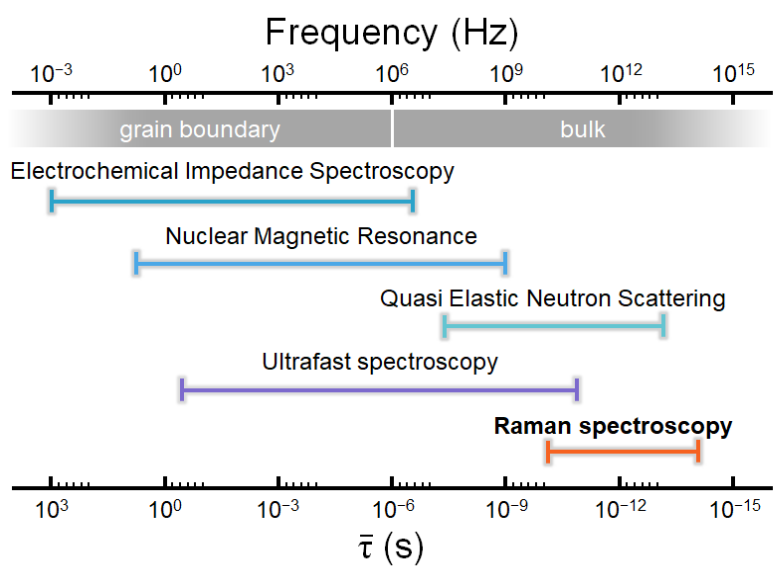

**Figure 1.** Comparison of commonly used techniques for detecting ion hopping and polyanion reorientation on different time scales.

In this work, we integrate *ab initio* molecular dynamics (AIMD) simulations with temperature-dependent Raman spectroscopy to investigate the reorientational dynamics of the $NO_2^-$ polyanion in two solid electrolytes, $Na_3ONO_2$ and $NaNO_2$. Our AIMD simulations reveal that the $NO_2^-$ polyanion in $Na_3ONO_2$ undergoes continuous 360° rotation, whereas in $NaNO_2$ the dynamics are restricted to rocking vibrations. Complementary Raman studies confirm these distinct behaviors. In $Na_3ONO_2$, the FWHM of internal vibrational modes increases linearly with temperature, indicating accelerated rotational dynamics; in $NaNO_2$, abrupt FWHM changes occur in association with its order–disorder phase transition. Based on the Raman results, we propose a general guideline for assessing the polyanion reorientation through FWHM variations of internal vibrational modes in Raman spectrum, a framework that can be extended to other polyanion-based solid electrolytes. By establishing a quantitative framework that links Raman spectral features to underlying anion reorientation dynamics, our approach offers researchers in solid-state ionics, energy storage, and computational materials science a versatile methodology for studying the interplay between lattice dynamics and ionic conductivity. Even though our findings for

Na$_3$ONO$_2$ indicate that NO$_2^-$ rotation does not directly enhance Na$^+$ diffusion, they provide critical insights into how polyanion dynamics can be decoupled from cation transport—a concept that can be extended to other material systems where such coupling may indeed improve conductivity. Thus, this work not only advances the understanding of polyanion reorientation in solid electrolytes but also bridges the gap between specialized spectroscopic techniques and broader applications in materials design.

**Results**

At room temperature, Na$_3$ONO$_2$ adopts a thermodynamically stable cubic phase with the space group $Pm\bar{3}m$,[32] as shown in Figure 2a. In this structure, [ONa$_6$] octahedra occupy the vertices of the cubic lattice and corner-linked *via* Na cations situated at the lattice edges. The incompatibility between the site symmetry ($m\bar{3}m$) and the point symmetry of the nitrite ion (2*mm*) entails an orientational disorder of the polyanion.[33] Each N and O atom in the NO$_2^-$ polyanion can occupy 6 and 24 crystallographically equivalent positions, respectively.[34] The X-ray diffraction (XRD) pattern of synthesized Na$_3$ONO$_2$ is presented in Figure S1. NaNO$_2$, the precursor for Na$_3$ONO$_2$ synthesis, is a well-known ferroelectric crystal that undergoes a primary phase transition from a ferroelectric phase to an anti-ferroelectric phase at 437.1 K, followed by a secondary phase transition to a paraelectric phase at 438.4 K.[35,36] Because the antiferroelectric phase exists only within a narrow temperature range (approximately 1 K), this study considers only the ferroelectric and paraelectric phases. In its ferroelectric phase (Figure 2b), NaNO$_2$ stabilizes in an ordered orthorhombic structure with space group *Im2m*, while the paraelectric phase (Figure 2c) features a disordered structure with space group *Immm*, reflecting an ordered-disordered phase transition.[37]

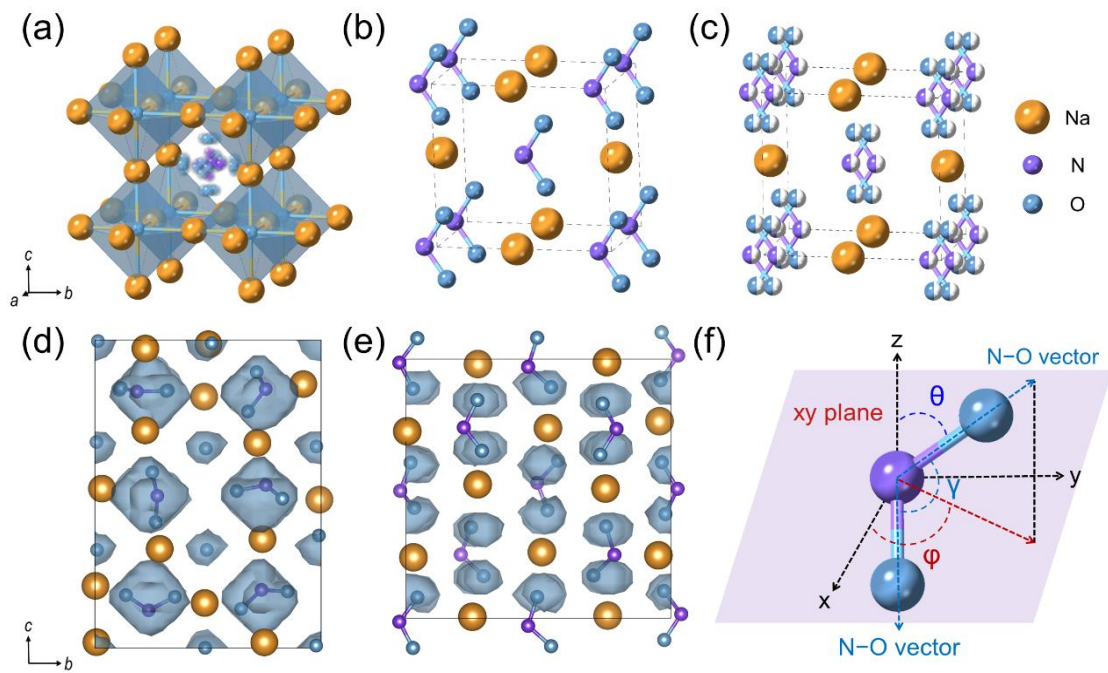

**Figure 2.** (a) Crystal structure of $Na_3ONO_2$ (Cubic, $Pm\bar{3}m$). (b, c) Crystal structures of the low-temperature orthorhombic (*Im2m*, left) and high-temperature orthorhombic (*Immm*, right) phases of $NaNO_2$. (d, e) The probability densities of O in (d) $Na_3ONO_2$ and (e) $NaNO_2$ at 500 K (blue isosurface), with an isosurface level of 0.001 1/bohr³. (f) Definition of the angles $\theta$, $\varphi$ and $\gamma$ in the reference frame of the crystal lattice, where $\theta$ is defined as the angle between the N-O vector and z axis, $\varphi$ corresponds to the angle between the x axis and the projection of the N-O vector in the xy plane, and $\gamma$ is the bond angle of the $NO_2^-$.

Thermogravimetric-differential scanning calorimetry (TG-DSC) curves in Figure S2 confirm that $Na_3ONO_2$ remains stable above room temperature without phase transitions. Conversely, the DSC curve for $NaNO_2$ shows a heat-absorption peak near 438 K, consistent with the ordered-disordered phase transition. The ionic transport properties of $Na_3ONO_2$ and $NaNO_2$ were further analyzed using temperature-dependent electrochemical impedance spectroscopy (EIS). Figures S3 and S4 show the Nyquist and Arrhenius plots of $Na_3ONO_2$ and $NaNO_2$ at various temperatures, respectively, indicating that the ionic conductivity of $Na_3ONO_2$ is significantly higher than that of $NaNO_2$. To explore the dynamics of the $NO_2^-$ polyanion in different structural environments, AIMD simulations were conducted at 325 K and 500 K for both $Na_3ONO_2$ and $NaNO_2$. The ion probability densities of $O^{2-}$ at 500 K show distinct behaviors (Figures 2d and 2e): in the cubic phase of $Na_3ONO_2$, O atoms in $NO_2^-$ rotate freely around N atoms (Figure 2d). In contrast, in the orthorhombic phase of $NaNO_2$, O atoms vibrate near their initial positions (Figure 2e). Additional ion probability densities for Na and O in both $Na_3ONO_2$ and $NaNO_2$ at 325 K and 500 K are presented in Figure S5, which corroborate these dynamic behaviors. To quantify these observations, the O ligands of $NO_2^-$ were mapped in spherical coordinates that defined in Figure 2f, where $\theta$ represents the angle between the N-O bond and the z-axis, $\varphi$ corresponds to the angle between the x-axis and the projection of the N-O vector in the xy-plane, and $\gamma$ represents the bond angle of the $NO_2^-$.

A detailed analysis of the AIMD results within the crystal lattice reference frame (Figure 3) further clarifies the different motions of the $NO_2^-$ polyanion. The preferential orientation of the $NO_2^-$ polyanion can be reflected by the spatial distribution of O atoms. Figures 3a and 3b show the two-dimensional (2D) projected probability distribution of O ligands in $NO_2^-$ in the defined spherical coordinates at 500 K. In $Na_3ONO_2$, fourteen distinct maxima (denoted as a, b, c, d, etc.) are observed for the two O atoms bonded to N, indicating facile $NO_2^-$ rotation occurs at this temperature. In contrast, $NaNO_2$ exhibits only five maxima for the O atoms (Figure 3b), with two prominent peaks (a, b) and three lower-intensity peaks (c, d, and e), suggesting that $NO_2^-$ polyanion in $NaNO_2$ undergoes limited rocking vibrations rather than full rotation.

To elucidate the origin of the different reorientation propensities of the $NO_2^-$ polyanion in the two structures, we calculated the Helmholtz free energy for the two O atoms bonded to N (Figures 3c and 3d). The Helmholtz free energy surface can be visualised as a "landscape", with "valleys" representing stable configurations (low free energy) and "peaks" corresponding to high energy, unstable states. As atoms explore this landscape, they naturally tend to settle into the valleys, driven by the tendency to minimize free energy—a balance between internal energy (favoring order) and entropy (favoring disorder). The migration or reorientation of ions in solid electrolytes is guided by the slopes and barriers of this free energy landscape. Therefore, the low free energy barriers (0.01−0.06 eV) of the reorientational dynamics allow these O ligands to easily rotate to adjacent minima. By analyzing this surface, we gain insights into the driving forces of the polyanion rotation in crystals. As shown in Figure 3c, the Helmholtz free energy surface of $NO_2^-$ in $Na_3ONO_2$ clearly reveals that the O ligands exhibit a very shallow and flat free energy surface. In contrast, Figure 3d shows that the Helmholtz free energy surface of the O ligands in $NaNO_2$ exhibit deep potential wells. The higher Helmholtz free energy barrier (0.03−0.27 eV) hinders the rotation of the $NO_2^-$ polyanion. The 2D projected probability distribution and Helmholtz free energy surface for the O ligands in $Na_3ONO_2$ and $NaNO_2$ at 325 K are presented in Figure S6. At 325 K, the maxima in the probability distribution of the oxygen ligands in $Na_3ONO_2$ decrease, and the free energy barrier increases slightly (0.01–0.08 eV), whereas in $NaNO_2$, only two probability maxima are observed with a free energy barrier of 0.34 eV, indicating that the reorientational dynamics of the $NO_2^-$ polyanion are negligible at this temperature.

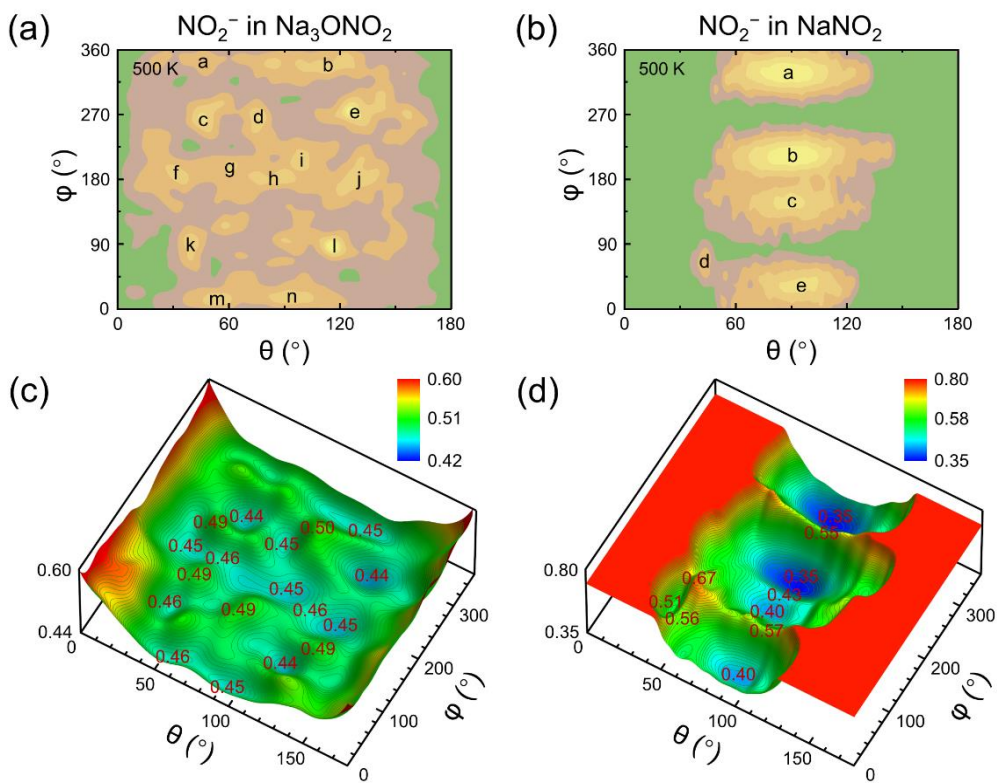

**Figure 3.** 2D projected probability distribution and Helmholtz free energy surface for O ligands in the $NO_2^-$ polyanion. (a, b) 2D projected probability distribution of O ligands in (a) $Na_3ONO_2$ and (b) $NaNO_2$. (c, d) Helmholtz free energy surface of O ligands in (c) $Na_3ONO_2$ and (d) $NaNO_2$ as a function of angle $\theta$ and $\varphi$. The Helmholtz free energy $A$ was computed as $A(\theta, \varphi) = -k_B T \ln[\rho(\theta, \varphi)]$, where $k_B$ is the Boltzmann constant, $T$ is temperature, and $\rho(\theta, \varphi)$ is the probability density distribution of the O ligands of the $NO_2^-$ polyanion from the AIMD simulations.

To quantify the rotation time of the $NO_2^-$ polyanion and compare its reorientation rates at different temperatures, we calculated the variations of angle $\theta$ (defined as the angle between the N-O vector and the z-axis in Figure 2f) with simulation time and determined the reorientation autocorrelation function, $C(t)$ (Figure 4). The angle $\theta$ is tracked as an indicator of the polyanion's reorientational dynamics, reflecting the spatial distribution of $NO_2^-$ polyanion relative to the z-axis.[20,21] Remarkable changes in $\theta$ over time are observed at both 325 K and 500 K in $Na_3ONO_2$ (Figures 4a and 4b), indicating facile and continuous rotation of the $NO_2^-$ polyanion during the simulations. At 325 K, $NO_2^-$ polyanion requires approximately 30–40 ps to complete a full rotation (360°), whereas at 500 K the rotation time decreased to 15–20 ps. Figure 4c shows that the reorientation autocorrelation function $C(t)$ of the $NO_2^-$ polyanion in $Na_3ONO_2$ decays rapidly at both temperatures, reaching zero in less than 5 ps. The rate decay of this function to zero reflects the reorientation rate, hence it confirms a significantly faster reorientation rate at 500 K compared to 325 K. In contrast, the variation of angle $\theta$ in $NaNO_2$ at 325 K and 500 K (Figures 4d and 4e) further suggests that the $NO_2^-$ polyanion undergoes limited rocking vibrations rather than rotations. The lower Helmholtz free energy barrier for the O ligands in

NaNO$_2$ at 500 K compared to 325 K implies an increase in the rocking amplitude at the higher temperature, which indicates greater disorder in the NO$_2^-$ orientation. The *C(t)* function for NaNO$_2$ remains close to one at 325 K and decays slowly at 500 K, demonstrating that the polyanion remains near its initial position at 325 K while exhibiting activated rocking vibrations at 500 K. Finally, the bond lengths and bond angles ($\gamma$) of the NO$_2^-$ polyanions in both Na$_3$ONO$_2$ and NaNO$_2$ remain stable over simulation time (Figures S7 and S8), although the amplitude of the stretching and bending vibrations increases at 500 K, confirming the overall structural stability with enhanced vibrational activity at elevated temperatures.

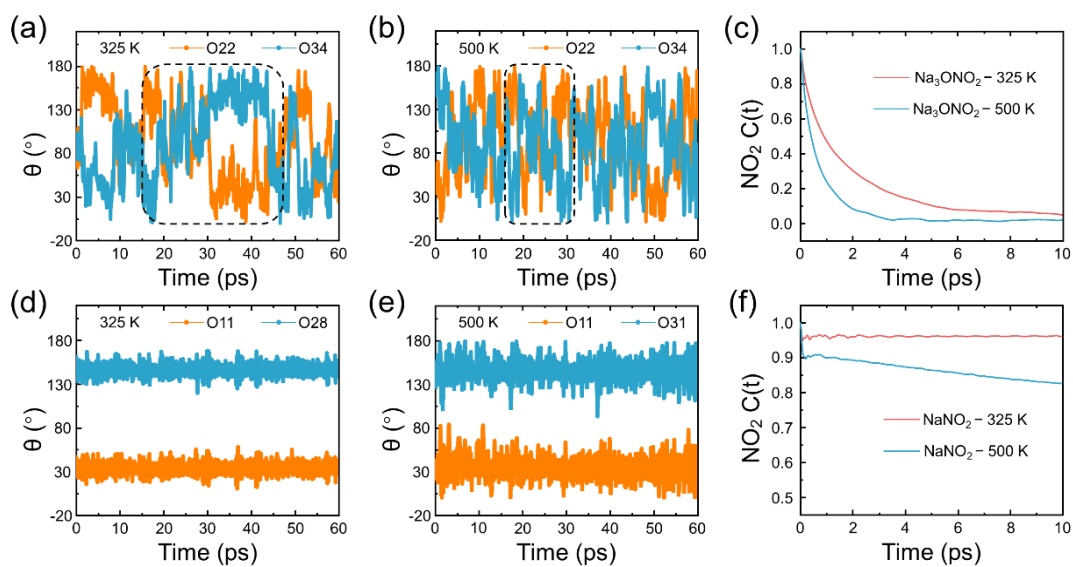

**Figure 4.** Reorientational dynamics of NO$_2^-$ polyanion in Na$_3$ONO$_2$ and NaNO$_2$ during the AIMD simulations. (a, b) The angle $\theta$ between the N-O vector and the z-axis in Na$_3$ONO$_2$ as a function of simulation time at (a) 325 K and (b) 500 K, respectively. (c) Reorientation autocorrelation functions of NO$_2^-$ polyanion in Na$_3$ONO$_2$. (d, e) The angle $\theta$ in NaNO$_2$ as a function of simulation time at (d) 325 K and (e) 500 K, respectively. (f) Reorientation autocorrelation functions of NO$_2^-$ polyanion in NaNO$_2$.

Based on the $C_{2v}$ symmetry of the free NO$_2^-$ polyanion, all nine optical phonon modes of Na$_3$ONO$_2$ and NaNO$_2$ are Raman active.[38] Figure 5a shows the room-temperature Raman spectrum of both compounds. The librational, rocking, and rotational modes of the NO$_2^-$ polyanion appear in the low-frequency range of 0−300 cm$^{-1}$.[39] The Raman spectrum of Na$_3$ONO$_2$ show that these modes merge to form a large broad peak. The internal vibrational modes (2A$_2$ + B$_1$) of NO$_2^-$ polyanion are observed in the high−frequency region above 700 cm$^{-1}$.[39] Figure 5b presents the three vibrational modes of the NO$_2^-$ polyanion. In Na$_3$ONO$_2$, the $v_1$ peak at 812 cm$^{-1}$ corresponds to the in-plane bending mode (A$_1$), and the $v_2$ peak at 1320 cm$^{-1}$ corresponds to phonon coupling modes (A$_1$ + B$_1$) arising from symmetric and antisymmetric stretching vibrations.[39] The peak at 1061 cm$^{-1}$ is attributed to the stretching vibration of impurity NaNO$_3$ produced during the reaction.[40,41] Similarly, in NaNO$_2$, the $v_1$ peak at 829 cm$^{-1}$ is assigned to an in-plane bending mode (A$_1$), the $v_2$ peak at 1327 cm$^{-1}$ corresponds to the symmetric stretching vibrational mode (A$_1$), and the $v_3$ peak at 1233 cm$^{-1}$ corresponds to the antisymmetric stretching vibrational mode (B$_1$).[39] Figure S9 shows the vibrational power spectrum obtained by calculation for each ion in Na$_3$ONO$_2$ and NaNO$_2$ at 325 K and 500 K, which are essentially identical to the features observed in the Raman spectroscopy. The vibrational peaks of N and O at high frequencies (20 < $f$ < 50 THz) correspond to the internal bending and stretching modes of the NO$_2^-$ polyanion.[39,42]

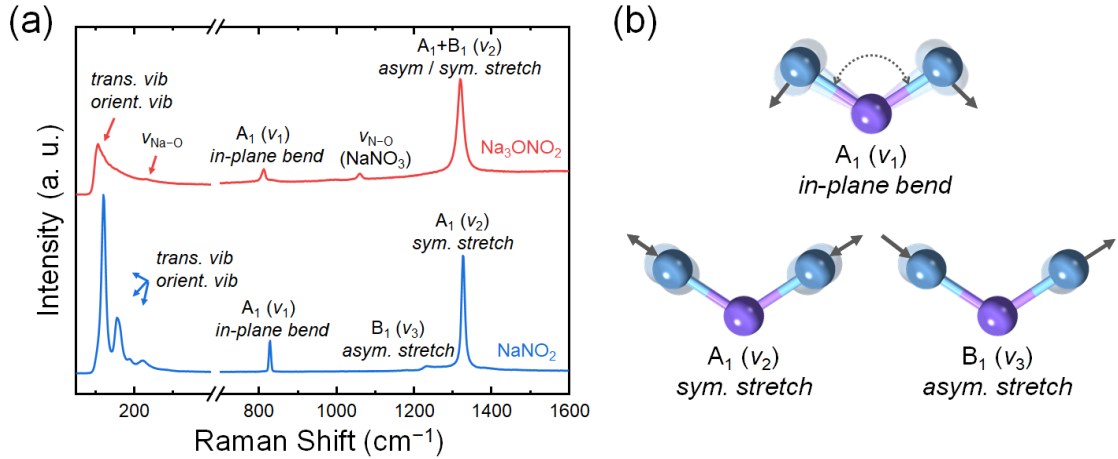

**Figure 5.** (a) Raman spectrum of Na$_3$ONO$_2$ and NaNO$_2$ at room temperature. (b) Three vibrational modes of the NO$_2^-$ polyanion: the in-plane bending mode (A$_1$), the symmetric stretching vibration mode (A$_1$) and the antisymmetric stretching vibration mode (B$_1$).

To experimentally elucidate the reorientational dynamics of NO$_2^-$ polyanion, we conducted temperature-dependent Raman spectroscopy on Na$_3$ONO$_2$ and NaNO$_2$. Figures 6a and 6c present the temperature-dependent Raman spectra of both compounds. The Raman spectrum of Na$_3$ONO$_2$ was measured from 293 K to 523 K with intervals of 20 K below 453 K and 10 K above, and that of NaNO$_2$ was measured from 296 K to 486 K at 10 K intervals. Notably, the Raman spectrum of NaNO$_2$ exhibits multiple modes at low frequencies degenerate as the temperature increases. Similarly, the antisymmetric and symmetric stretching vibrational modes at 1233 cm$^{-1}$ and 1327 cm$^{-1}$ gradually merge with increasing temperature. This is also observed in the vibration power spectrum in Figure S9.

To facilitate a comprehensive analysis of the changes in the vibrational modes, we performed Gauss-Lorentz fitting on the $v_1$ and $v_2$ peaks of the Raman spectrum to obtain the temperature dependence of FWHM.[43] The anisotropic broadening of the internal vibrational modes of polyanion consists of vibrational broadening $\Gamma_v$ and reorientational broadening $\Gamma_R$[30]:

$$\Gamma_{aniso} = \Gamma_v + \Gamma_R \qquad (1)$$

where $\Gamma_{aniso}$ is the FWHM of the anisotropy spectrum. Assuming that the vibrational and reorientational motions are uncorrelated, and neglecting any potential effects from dipole-dipole coupling and collision-induced effects, the reorientation time is inversely proportional to the reorientational broadening $\Gamma_R$.[30] Previous studies have shown that the vibrational broadening $\Gamma_v$ varies less with temperature compared to the reorientational broadening $\Gamma_R$ and can be approximately neglected.[30,31] Therefore, the temperature-dependent FWHM for polyanions' internal vibrational modes can be regarded as a measure of the reorientational broadening $\Gamma_R$. This implies that an increase in the FWHM of the internal vibrational modes indicates an acceleration of the polyanion's reorientational motion.

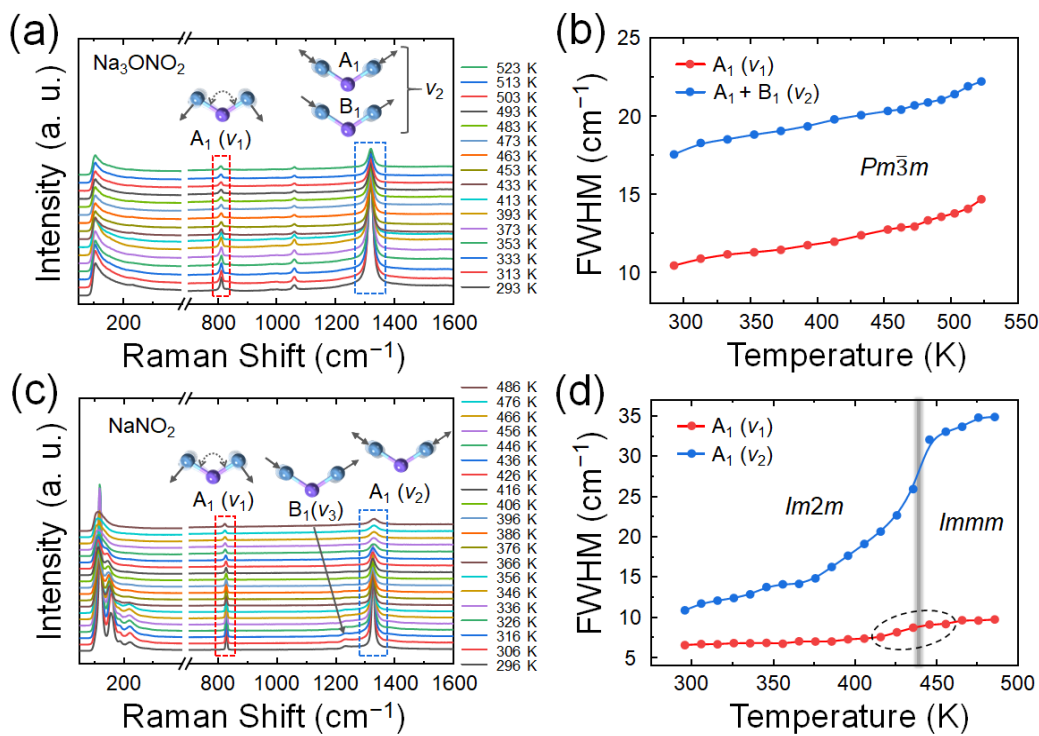

**Figure 6.** Temperature-dependent Raman spectrum of (a) $Na_3ONO_2$ and (c) $NaNO_2$. Variation of FWHM of (b) vibrational modes of $Na_3ONO_2$ and (d) vibrational modes of $NaNO_2$ with experimental temperature.

The fitting results of FWHM are presented in Figures 6b and 6d. In $Na_3ONO_2$, both internal vibrational modes exhibit an approximately linear increase in FWHM, indicating that the rotation of $NO_2^-$ polyanion accelerates with increasing temperature and leads to greater disorder. In contrast, $NaNO_2$ shows different trends for the two internal vibrational modes, in general agreement with previous studies by Rysiakiewicz-Pasek et al.[38] and Von Der Lieth et al.[44] In $NaNO_2$, the FWHM of the symmetric stretching mode increases significantly before and during the phase transition, and stabilizes after the phase transition, likely due to the coupling with the antisymmetric stretching mode, which leads to an anomalous FWHM increase. Conversely, the FWHM of the in-plane bending mode remains constant before the phase transition and increases by only ~2 cm$^{-1}$ during the transition. The abrupt change in FWHM near the phase transition temperature indicates that $NaNO_2$ undergoes an ordered-disordered phase transition, resulting in reorientational motion of the $NO_2^-$ polyanion. We also fitted the temperature-dependent frequency shifts of the internal vibrational modes of the $NO_2^-$ polyanion in both structures. As shown in Figure 7a, the stretching vibrational mode in $Na_3ONO_2$ remains nearly unchanged, while the bending vibrational mode exhibits a slight phonon softening with increasing temperature. This suggests that an increase in the orientational freedom of the $NO_2^-$ polyanion more significantly affects the bending vibrational mode. In contrast, Figure 7b shows the stretching and bending vibrational modes in $NaNO_2$ undergo significant phonon hardening and softening, respectively, near the phase transition temperature. The stretching vibrational hardening arises from the enhanced local coupling (increased rigidity) of the N−O bonds in the disordered phase, whereas the bending vibrational softening reflects a reduction in the bond angle restoring force due to increased orientational disorder. Both reflect the increased orientational freedom and lattice symmetry resulting from the ordered−disordered phase transition in $NaNO_2$.

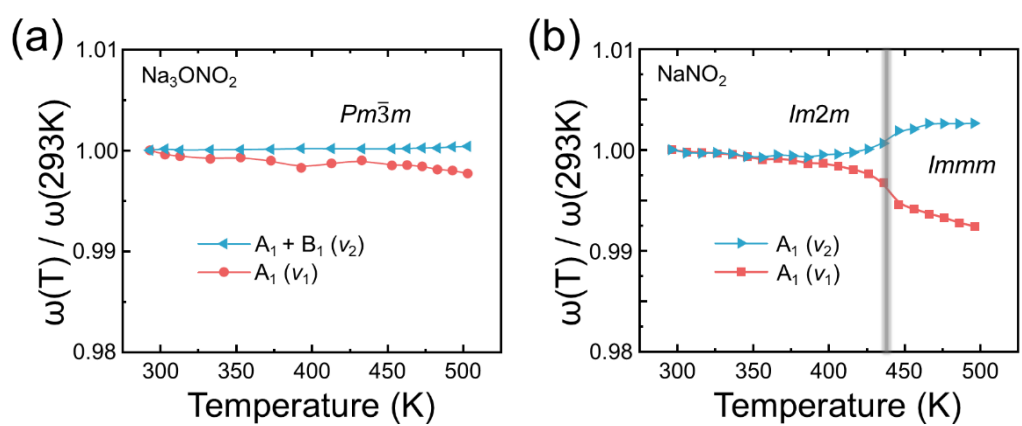

**Figure 7.** Temperature dependence of frequency shifts of the Raman modes in (a) $Na_3ONO_2$ and (b) $NaNO_2$.

**Discussion**

Based on the temperature-dependent Raman spectrum of $Na_3ONO_2$ and $NaNO_2$, we propose a general guideline for assessing the reorientation motion of the polyanion. The full width at half maximum (FWHM) of the polyanions' internal vibrational modes remains constant before the phase transition and exhibits a distinct jump during the phase transition. This behavior indicates that the structure undergoes an ordered-disordered phase transition, causing the polyanion changing from static to rocking vibrations or even to complete rotational motion. In contrast, a significant and continuous linear increase in the FWHM with rising temperature indicates that the polyanion is undergoing continuous and complete rotation. Figure 8 schematically illustrates how Raman spectroscopy characterizes the reorientational dynamics of polyanions.

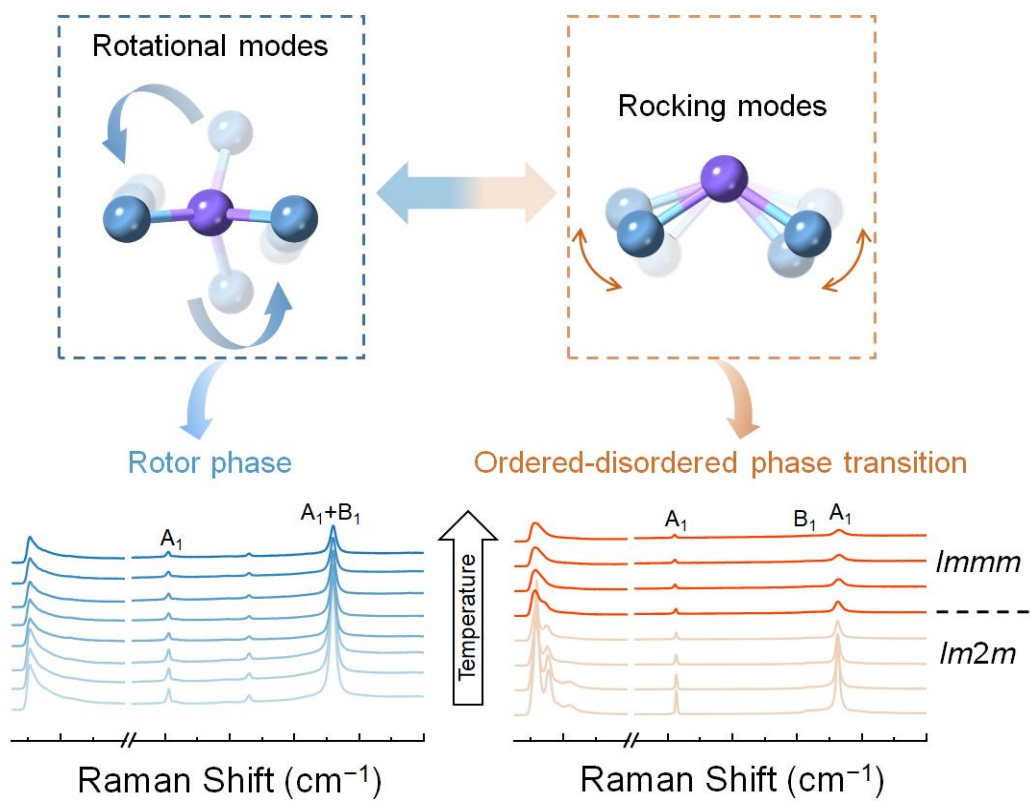

**Figure 8.** Schematic representation of Raman spectroscopy characterizing polyanion dynamics.

We further investigated the interaction between polyanion reorientation and cation translational motions of the two structures. Figure S10 shows the power spectrum of $Na^+$ vibrations and $NO_2^-$ librations in $Na_3ONO_2$ and $NaNO_2$ (less than 10 THz). The power spectrum of the $Na^+$ and $NO_2^-$ exhibit a considerable overlap in the low-frequency region, suggesting that their vibrational frequencies are similar. However, the mean squared displacement (MSD) data (Figure S11) reveal that the reorientation of the polyanion is decoupled from cation migration. The $N^{3+}$ and $O^{2-}$ atoms of the $NO_2^-$ polyanion in $Na_3ONO_2$ exhibit significant displacements, whereas the displacement of $Na^+$ remains relatively minor. This finding suggests that the rotation of the $NO_2^-$ polyanion does not enhance cation diffusion, a conclusion that is consistent with the observed local $Na^+$ probability densities (Figure S5). At both 325 K and 500 K, the $Na^+$ trajectories demonstrate local $Na^+$ vibration with no long-range transport in $Na_3ONO_2$. In $NaNO_2$, the MSD for each ion at 325 K is approximately equal to zero over the simulation time. In contrast, the MSDs for $Na^+$, $N^{3+}$, and $O^{2-}$ all exhibit a gradual increase at 500 K. This behavior implies that elevated temperatures enhance $Na^+$ vibrations rather than coupling with the $NO_2^-$ rocking vibrations, which is consistent with the inherently low ionic conductivity of $NaNO_2$.

To further explore the decoupling mechanism between $NO_2^-$ polyanion rotation and $Na^+$ migration in $Na_3ONO_2$, we performed AIMD simulations on $Na_3OBH_4$, a structural analog that shares the same O-based framework and has a comparable unit cell volume.[45] As demonstrated in Figure S12, $Na_3OBH_4$ exhibits rapid $BH_4^-$ polyanion rotation at both 300 K and 500 K; however, this dynamic rotation similarly fails to enhance Na$^+$ diffusion. Notably, other Na$_3$OX (X = Cl, Br, I) solid electrolytes also demonstrate low ionic conductivities.[45,46] These observations collectively attribute the poor ionic transport in this material family to the inherent rigidity of their O-based frameworks. The O$^{2-}$ at lattice vertices form a structurally inflexible matrix that impedes local structural relaxation, thereby maintaining elevated transition-state energy barriers for Na$^+$ migration. Consequently, Na$^+$ must overcome intensified electrostatic repulsion and lattice distortion resistance during hopping processes, which fundamentally disrupts the formation of continuous migration channels and precludes dynamic coupling with polyanion rotation at lattice centers. This rigid O-based framework acts as both a structural confinement for Na$^+$ migration and a dynamic cage for polyanion rotation, suppressing their coupling interactions. This suppression underscores the pivotal role of lattice softening in designing high-performance ionic conductors. In contrast, solid electrolytes that exhibit the "paddle-wheel effect", such as Li$_2$OHCl[19,20] and Na$_3$PX$_4$ (X = S, Se)[13], possess more flexible lattices. The larger ionic radii and weaker bonding strength of Cl$^-$ or S$^{2-}$/Se$^{2-}$ confer lattice flexibility, enabling dynamic coupling between OH$^-$ or PX$_4^{3-}$ rotation and cation migration. Furthermore, in Na$_3$PX$_4$ systems, the high polarizability of S/Se atoms induces significant dipole moment variations during PX$_4^{3-}$ rotation, generating dynamic electric fields that guide Na$^+$ migration along low-energy pathways. Conversely, the NO$_2^-$ group in Na$_3$ONO$_2$ demonstrates limited polarization capability due to the strong covalency of N–O bonds and the high electronegativity of O atoms, resulting in insufficient modulation of local electric fields around Na$^+$ during rotation.

Understanding the rotational dynamics of polyanions is critical, as their reorientation behavior can significantly modulate ionic migration energy barriers. Our study establishes a novel methodology for characterizing polyanion reorientation dynamics in solid electrolytes using temperature-dependent Raman spectroscopy, complemented AIMD simulations to elucidate the interaction between polyanion rotation and cation migration. This integrated approach provides atomic-scale insights into the microscopic mechanisms governing ionic conductivity regulation. Although NO$_2^-$ polyanion rotation in Na$_3$ONO$_2$ does not directly facilitate Na$^+$ migration, the developed theoretical framework for assessing polyanion reorientation dynamics holds broader implications for diverse material systems. In the solid electrolytes that exhibit the "paddle-wheel effect", such as Li$_2$OHCl and Na$_3$PX$_4$ (X = S, Se), the rotational dynamics of OH$^-$ or PS$_4^{3-}$/PSe$_4^{3-}$ polyanions can be experimentally probed through Raman spectroscopy analysis. Specifically, thermal broadening of O–H stretching modes (in Li$_2$OHCl) or P–S/P–Se stretching/bending vibrational modes (in Na$_3$PX$_4$) provides direct spectroscopic features of polyanion rotation. These experimental observations can be synergistically combined with AIMD simulations to investigate the dynamic coupling between polyanion rotation and cation migration pathways. This theoretical framework extends beyond the confines of specific material systems, establishing generalized design principles for the rational optimization of solid-state ionic conductors.

**Materials and Methods**

**Material Synthesis.** Na$_3$ONO$_2$ was synthesized using a classical solid-state method, with all procedures conducted under an Ar atmosphere. Na$_2$O (Alfa Aesar, 80%) and NaNO$_2$ (Macklin, 99.99%) were used as precursors. Stoichiometric amounts of the precursors were thoroughly mixed and ground in an agate mortar for 30 minutes. The resulting mixture was then pressed into a pellet with a diameter of 12mm at 300 MPa using a stainless-steel die. The pellet was sealed in a Pyrex glass tube under vacuum pressure below $2 \times 10^{-3}$ Pa and sintered at 310 °C for 12 hours, following a heating rate of 3 °C min$^{-1}$. After sintering, the pellet was allowed to cool naturally to room temperature.

**X-Ray Diffraction (XRD).** Measurement was conducted on a Bruker D8 ADVANCE diffractometer with Cu Kα radiation at a current of 40 mA and voltage of 40 kV, scanning from 15° to 90° at a rate of 1° min$^{-1}$. The sample was placed on a Quartz glass sample stage and sealed by Kapton film under Ar atmosphere to prevent exposure to oxygen and moisture.

**Thermogravimetric Analysis and Differential Scanning Calorimetry (TG–DSC).** Samples were sealed inside a hermetic alumina crucible. TG–DSC measurements were performed using a NETZSCH STA 449 FA *Jupiter* instrument over a temperature range of 25–300 °C at a temperature scan rate of 5 °C min$^{-1}$.

**Electrochemical impedance spectroscopy (EIS).** The electrochemical impedance spectroscopy (EIS) measurements were measured using a VIONIC electrochemical workstation at frequencies ranging from 1 Hz to 10 MHz with an amplitude of 50 mV. The sample powders were cold-pressed into pellets (diameter = 12 mm, thickness ~ 0.6 mm) under 300 MPa using a polyetheretherketone (PEEK) insulated die. Two stainless-steel rods with a diameter of 12 mm were clamped on both sides of the sample as current collectors. The ionic conductivity (σ) was then calculated from the following formula:

$$\sigma = \frac{L}{RS} \qquad (2)$$

where $L$ is the thickness of the samples after pelleting, $R$ is the resistance of the measured samples, and $S$ is the area of the samples after pelleting.

**Raman Spectroscopy.** Raman spectrum is recorded using a Renishaw inVia Raman spectrometer with a 532 nm laser, and the laser output power was maintained between 20 and 50 mW. The incident light is made perpendicular to the scattering plane using a Grand Taylor prism, and the scattered light is made parallel to the scattering plane by rotating the analyzer. The direction of polarization was verified by measuring the receding bias ratio of the single-crystal quartz. High–temperature Raman spectroscopy measurements were performed using a LINKAM TS1500V stage at a heating rate of 5 °C min$^{-1}$ from 20 to 250 °C. N$_2$ was continuously and slowly fed into the LINKAM stage to protect the samples from oxygen and humidity during the measurements.

**Computational Details.** Geometry optimization was carried out by density functional theory (DFT)[47] calculations using the Vienna *ab initio* simulation package (VASP)[48] with the projector augmented-wave (PAW)[49] method. The generalized gradient approximation (GGA)[50] function parameterized by Perdew-Burke-Ernzerhof (PBE)[50] was used to describe the exchange correlation potential. A cutoff energy of 450 eV was used in the calculations.

For structural optimization calculations, the Brillouin zone was sampled by using a 3 × 3 × 3 Γ-centered k-point grid. The total energy and the force on each atom were converged to within 10$^{-5}$ eV and 0.01 eV Å$^{-1}$, respectively.

AIMD simulations were carried out at a constant pressure within the canonical (NVT) ensemble using a Nose thermostat[51] to maintain a constant temperature. The volume of the unit cell was fixed at that of the fully relaxed structure. Na$_3$ONO$_2$ was simulated using a 3 × 3 × 2 repetitive supercells with cubic ($Pm\bar{3}m$) structure, while the low temperature phase NaNO$_2$ was simulated using a 3 × 3 × 3 repetitive supercells with orthorhombic ($Im2m$) structure, and the high temperature phase NaNO$_2$ was simulated using 2 × 2 × 2 repetitive supercells with orthorhombic ($Immm$) structure. Na$_3$OBH$_4$ was simulated using a 3 × 2 × 2 repetitive supercells with cubic ($Pm\bar{3}m$) structure. All simulations were thermalized to 300 K, 325 K and 500 K, equilibrated for 5 ps, and then equilibrated at the desired temperature for 60 ps. To keep the computational time reasonable for the relatively large unit cell, we only performed integration in reciprocal space at the Γ-point.

Na$^+$-ion and O$_2^-$-ion probability densities were calculated from the atomic trajectories. The diffusion coefficient is defined as

$$D = \lim_{t \to \infty} \left[\frac{1}{2dt} |\vec{r}_i(t) - \vec{r}_i(0)|^2\right] \quad (3)$$

where *d* is the dimension of the lattice on which the diffusion takes place, and *t* is the elapsed time. The average mean square displacement is the averaged displacement of each atom over time *t*:

$$\langle|\vec{r}(t) - \vec{r}(0)|^2\rangle = \frac{1}{N}\sum_{i=1}^{N} \langle|\vec{r}_i(t + t_0) - \vec{r}_i(t_0)|^2\rangle \quad (4)$$

where $\vec{r}_i(t)$ is the displacement of the *i*th ion at time *t*, and $t_0$ is the initial time. *D* is obtained by linear fitting to the dependence of average-mean-square displacement over 2*dt*.

We also computed the normalized 2D probability density distribution $\rho_{\theta,\varphi}^{2D}$ as a function of $\theta$ and $\varphi$ for the O ligands of NO$_2^-$ polyanion from the AIMD simulations. $\theta$ represents the angle between a selected N−O bond and the z axis, and $\varphi$ corresponds to the angle between the x axis and the projection of the same N−O vector in the *xy* plane. On the basis of this, the Helmholtz free energy surface of the O ligands was computed via

$$A(\theta, \varphi) = -k_B T \ln \rho_{\theta,\varphi}^{2D} \qquad (5)$$

where $k_B$ is the Boltzmann constant, and $T$ is temperature.

The reorientation autocorrelation function $C(t)$[52] is defined as

$$C(t) = \langle u(t) \cdot u(t=0) \rangle \qquad (6)$$

where $u(t)$ is a unit vector from the center of mass (*i.e.*, the position of the central N atom) of the polyanion to a constituent anion (O atom) at time $t$. $\langle\ \rangle$ denotes the averaged value across all anions of the same species. The rate at which this function decays to 0 reflects the reorientation rate of that species.

The power spectrum for each ion vibration was calculated *via* the Fourier transform of the velocity autocorrelation function, $\langle x(t') \cdot x(t'+t) \rangle$. $x$ is the linear velocity ($v_k$) of the cation or angular velocity ($w_k$) of the anion. $w_k$ was calculated via

$$w_k = \frac{r_k \times v_k}{r_k^2} \qquad (7)$$

Here, $r_k$ is the position vector for each anion atom relative to the central of mass of the polyanion.

## Acknowledgments

The authors of this work gratefully appreciate the financial support provided by the National Natural Science Foundation of China (Nos. 12475344, 12375343)